\documentstyle[PASJadd,psfig]{PASJ95}

\newcommand{\vol}{52}
\newcommand{\no}{2}
\newcommand{\lett}{}
\newcommand{\spage}{}
\newcommand{\rdate}{1999 August 31}
\newcommand{\adate}{1999 December 15}

\begin{document}
\title{Broad-Band X-Ray Study of a Transient Pulsar RX~J0059.2$-$7138}
\author{Makoto {\sc Kohno}, Jun {\sc Yokogawa}, 
and Katsuji {\sc Koyama}\thanks{CREST, Japan Science and Technology
Corporation (JST), 4-1-8 Honmachi, Kawaguchi, Saitama 332-0012.} \\ 
{\it Department of Physics, Graduate School of Science, Kyoto University, 
Sakyo-ku, Kyoto 606-8502} \\
{\it E-mail(MK): kohno@cr.scphys.kyoto-u.ac.jp}}

\abst{
We report on the results of the ASCA and ROSAT observations on
RX~J0059.2$-$7138, a transient X-ray pulsar in the Small Magellanic
Cloud. 
The barycentric pulse period has been precisely determined to be $2.763221 \pm
0.000004$~s.  
The pulse shape is almost identical in all of the energy bands. 
The pulse fraction increases with the photon energy below $\sim 2$~keV,
while it is nearly 
constant at $\sim 37$\%  above $\sim 2$~keV.
The X-ray spectrum has been found to consist of two components.  One is
dominant above  2~keV, and exhibits sinusoidal pulsations.  This
component is well described by a typical model  found in many X-ray
binary pulsars,  a power-law  of photon index 0.4  
with an exponential cut-off at  6.5~keV.
The other is dominant below 1~keV and shows
no significant pulsation. This component is represented by either a
broken power-law  with photon indices of 2.6 and 5.1 below and above a
break energy of 0.9~keV,  or a metal-poor thin-thermal plasma with a
temperature of 0.37~keV. 
The phase-averaged luminosity is $\sim 10^{38}$~erg~s$^{-1}$
(0.1--10.0~keV) for both components.
A hint of oxygen over-abundance is found in  the absorbing column,
possibly due to circumstellar gas ejected from an evolved
companion. 
}

\kword{circumstellar matter --- pulsars: individual (RX~J0059.2$-$7138)
--- stars: Be --- stars: neutron --- X-rays: binaries}

\maketitle
\thispagestyle{headings}

\section{Introduction}

Accretion-powered X-ray pulsars are binary systems consisting of a
neutron star and a stellar companion (here,  X-ray binary pulsars or
XBPs in short).  
The gravitational energy of  accreting matter is converted to X-ray
radiation, hence  its luminosity is generally  variable, depending on the
mass-accretion rate.
Transient behavior is also observed from many XBPs,  most of which
have a Be star companion with an eccentric orbit (e.g.\ Stella et al.\
1982; Bildsten et al.\ 1997).

XBPs have been observed mainly in the hard X-ray band ($\sim 2$--40~keV) 
with non-imaging satellites (e.g.\ Nagase 1989).  
Their spectra are generally  described  by a power-law with a
high-energy exponential cut-off (hereafter, ECUT power-law) 
with a photon index of $\Gamma \sim 1$ and a cut-off energy at around 
10--20~keV.
A fluorescent emission line at 6.4~keV  from neutral iron atoms has been 
observed in many bright XBPs, which serves as a probe of the
circumstellar matter of the binary system.  
In the soft X-ray band (below $\sim 2$~keV), however,  their spectra
have not been well studied due to a large interstellar absorption,
because most of the XBPs are located in the galactic plane (e.g.\
Bildsten et al.\ 1997).  

In order to study soft X-rays from  XBPs, therefore,  the Magellanic
Cloud sources  have great advantages, owing to the low interstellar
absorption and  well-calibrated distances.  
Woo et al.\ (1995, 1996) made broad-band spectroscopic
studies in the 0.1--40~keV band of SMC X-1 and LMC X-4, respectively,
and found a ``soft excess'' below $\sim 2$~keV, in addition to the usual
hard spectrum given by an ECUT power-law model. 
Such a soft excess was also discovered from a transient XBP 
XTE~J0111.2$-$7317 in the Small Magellanic Cloud (SMC), 
by an ASCA observation in the 0.5--10.0~keV band (Yokogawa et al.\ 2000). 
However, since such a broad-band spectroscopy has been  performed on
only a few XBPs, whether or not  the soft component is common among XBPs 
is still unclear, especially for transient XBPs. 

Hughes (1994) serendipitously discovered a new transient XBP, 
RX~J0059.2$-$7138, in the SMC with the soft
X-ray band observation of ROSAT,  and found  that the spectrum is
unusually soft, which is composed of a blackbody with a temperature  of
$kT \sim 35$~eV  and a steep power-law  with a photon index of $\Gamma
\sim 2.4$.  
A proposed candidate for the optical counterpart was revealed to be
a Be star by Southwell and Charles (1996). 

In order to examine the nature  of the  unusually soft spectrum of
RX~J0059.2$-$7138, observations in the higher energy band are
essential. RX~J0059.2$-$7138 was also detected in a simultaneous
observation of the SNR E~0102$-$72.2 with the ASCA satellite, which is
sensitive  in the $\sim 0.7$--10.0~keV band.  A preliminary short report
on the ASCA results can be found in Kylafis (1996). 

In this paper, we combine the ROSAT and ASCA data,  and perform  a
broad-band timing  spectroscopy, covering $\sim 0.1$--10.0~keV.  
We show that the spectrum of  RX~J0059.2$-$7138 has a hard component,
which resembles the spectra  of usual  XBPs well,  and that the
pulsations are due to the hard component.  In addition,  a
non-pulsating soft component is present in the spectrum.  
We assume  the distance to RX~J0059.2$-$7138 to be 60~kpc, 
the nominal value to the SMC (Mathewson 1985). 

\section{Observations}
 
The transient XBP RX~J0059.2$-$7138 was serendipitously detected on May
12--13 in 1993, in the simultaneous ASCA and ROSAT observations of
E~0102$-$72.2, the brightest supernova remnant in the SMC. ASCA and
ROSAT observations spanned about 99~ks (MJD 49119.38--MJD 49120.52) and
20~ks (MJD 49119.95--MJD 49120.18), respectively.

ASCA carries four X-ray Telescopes (XRT)  with two Gas Imaging
Spectrometers (GIS~2 and GIS~3)  and  two Solid-state Imaging
Spectrometers (SIS~0 and SIS~1)  on the focal planes (Tanaka et al.\ 1994;
Serlemitsos et al.\ 1995; Ohashi et al.\ 1996; Burke et al.\ 1994). 
Since RX~J0059.2$-$7138 was outside of the SISs' field of view and was
located near the calibration source of GIS~3,  we used only the GIS~2
data in this paper.  
The GISs were operated in  the normal PH mode with a time resolution of
0.0625~s (high bit rate) or 0.5~s (medium bit rate).  
We rejected any data obtained when the satellite was in the SAA (South
Atlantic Anomaly) region, when the elevation angle of the Earth limb was 
less than $5^\circ$ or when the cut-off rigidity was less 
than 4~GV,  and also removed particle events using the rise-time
discrimination method.  
The total exposure time of GIS~2 was $\sim $35~ks after screening.

ROSAT  carries a soft X-ray telescope with a High Resolution Imager (HRI)
and a Position Sensitive Proportional Counter (PSPC) with one of the two
on the focal plane (Tr\"{u}mper 1983).
In this  observation,  PSPC-B was on the focus and operated in the
pointing mode. We took the screened data from the HEASARC archive, of
which the total exposure time was $\sim 5$~ks.  

\section{Analyses and Results}

For the GIS data,  X-ray photons were collected from  an elliptical
region around  RX~J0059.2$-$7138 (figure 1). 
The background data  were extracted from  off-source areas  on the edge
of GIS~2  at the same off-axis angle as the source (figure 1),
because  both the diffuse X-rays and non-X-ray background depend on the
off-axis angle. 

\begin{figure}[hbtp]
 \psfig{file=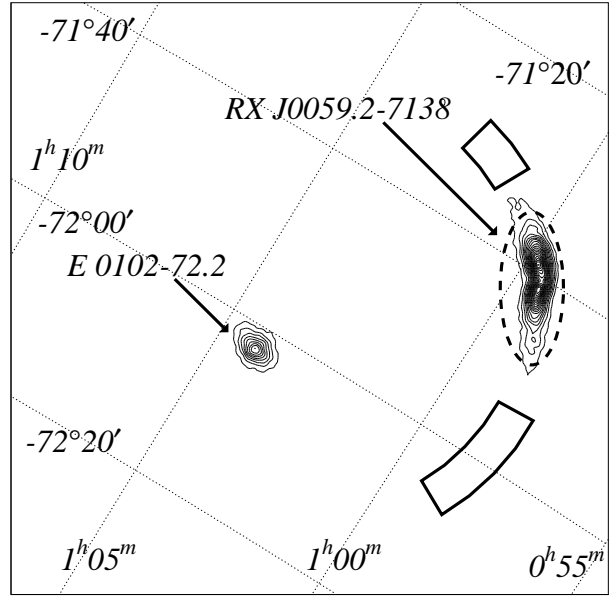,width=8cm}
 \caption{ASCA GIS~2 image. 
 The sources detected at the center and the edge are  the SNR
 E~0102$-$72.2 and RX~J0059.2$-$7138, respectively. 
 The photons of RX~J0059.2$-$7138 were spreaded by the point-spread function
 at the detector edge region.
 The source and background regions used in the analyses are enclosed by
 the dashed  and solid lines, respectively.} 
\end{figure}

As for the ROSAT PSPC, we selected  a circular region of $3'$-radius
around the source,  while the data from an annulus of inner and outer
radii $3'$ and $6'$ were  used as the background. 
Since a super-soft source (SSS) 1E~0056.8$-$7154 was located at $4'$
northwest of RX~J0059.2$-$7138  and the X-ray flux of this SSS was
fairly large in the ROSAT band below $\sim 0.5$~keV, we excluded  a
circular region of a $3'$-radius  around this SSS  from the analyses of
the ROSAT PSPC.

\subsection{Timing Analysis}

In total, the numbers of counts  from the source regions are 21847  in
the 0.7--10.0~keV band (ASCA GIS~2) and  23250  in the  0.07--2.4~keV
band (ROSAT PSPC). 
After the barycentric correction of arrival time for each event, we
carried out timing analyses. 
 
The pulse period was already determined to be 2.7632$\pm0.0002$~s  with
the ROSAT data alone (Hughes 1994).  We used the ASCA data to
obtain a more precise  period, because the
duration of the ASCA observation is nearly five-times  longer than the
ROSAT observation. 
We performed epoch-folding on all of the ASCA GIS~2 data, and  determined
a trial pulse period to be 2.76322~s. 
We then divided the ASCA observation into three segments,  each having a
duration of $\sim 33$~ks,  and folded each of the light curves with a
trial period of 2.76322~s. 

To the folded pulse profiles of the first and the last segments, we made
a cross correlation,  and determined the average apparent barycentric
pulse period during the observation to be 2.763221$\pm$0.000004~s.  This
value  is consistent with, and  more precise than,  the previous  result
of 2.7632$\pm0.0002$~s (Hughes 1994).  

The cross correlation between the middle  and the first/last segments
shows the upper limit of the phase difference to be  $\sim 0.04$. 
Based on the assumption of a constant period change during the
observation, this upper limit of the 
phase difference is converted  to the upper limit of the period
derivative of  $|\dot{p}| < 6\times10^{-10}$~s~s$^{-1}$.  

Figure 2 shows the pulse profiles in the six energy bands 
(0.07--0.4~keV, 0.4--1.0~keV, and 1.0--2.4~keV in ROSAT PSPC, and
0.7--2.0~keV, 2.0--4.0~keV, and 4.0--10.0~keV in ASCA GIS) folded with
a pulse period of 2.763221~s.
The pulse fractions, defined as $(I_{\rm max} - I_{\rm min}) /
2I_{\rm average}$, 
  in these energy bands are  0.07$\pm$0.06, 0.14$\pm$0.04, 0.26$\pm$0.04 
(0.07--0.4~keV, 0.4--1.0~keV, and 1.0--2.4~keV in ROSAT PSPC) and
0.24$\pm$0.05, 0.37$\pm$0.05, and 0.36$\pm$0.05  (0.7--2.0~keV, 
2.0--4.0~keV, and 4.0--10.0~keV in ASCA GIS),  respectively.  
The pulse shape is nearly sinusoidal and independent of the X-ray energy.
The apparent decrease of the pulse fraction with decreasing energy,
which was already found in the ROSAT data (Hughes 1994),  is also seen
in the ASCA data below $\sim 2$~keV. 

On a longer timescale (50--100~s), Hughes (1994) reported a flickering
variability of about 30\% of the mean rate in the ROSAT data. Such
variability was also seen in the ASCA data with no energy dependence,
while no larger variability, like a burst, was detected.

\begin{figure}[ht]
\psfig{file=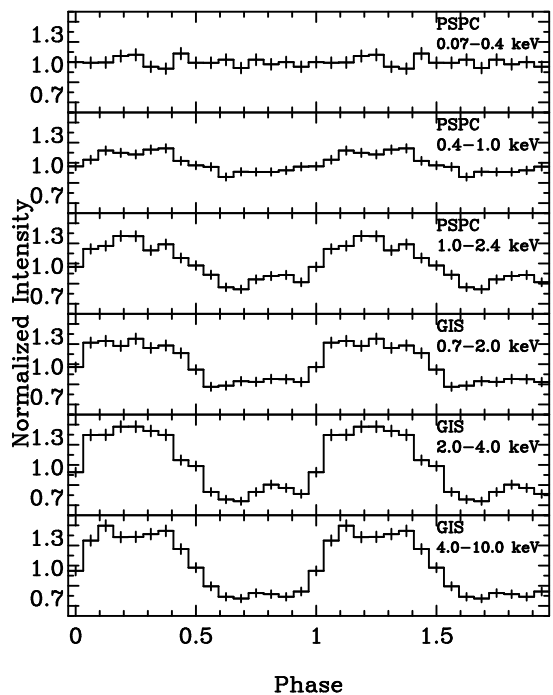,width=8cm}
\caption{Normalized intensities of the folded light curves. The background
 levels are $\sim 1$\% and $\sim 10$\% of the average, in the PSPC
 data and in the GIS~2 data, respectively.}
\end{figure}

\subsection{Spectral Analysis}

Figure 3 shows the phase-averaged spectra after background
subtraction.  Since the pulse profile is essentially energy independent 
while the pulse  fraction increases with energy below
$\sim 2$~keV,  we can naturally assume that  the spectrum of
RX~J0059.2$-$7138 consists of two components: a hard component dominant 
above $\sim 2$~keV, which has an energy-independent pulsation, and a
soft component dominant below  $\sim 2$~keV with no pulsation. 

With this working hypothesis, we first determined the hard component
using the ASCA spectrum  above $\sim 2$~keV.  Since no significant
emission line was  detected,  we fitted the spectrum with a single
power-law model. 
However, because systematic negative residuals remained above $\sim 7$~keV,
we introduced a high-energy exponential cut-off to the single
power-law model (ECUT power-law) and fitted the spectrum again.
We then obtained an acceptable fit to an ECUT power-law model with a photon 
index of  $\Gamma \sim 0.4$, a cut-off energy of $E_{\rm c} \sim
6.6$~keV and a folding energy of $E_{\rm f} \sim 8.8$~keV.

\begin{figure}[htp]
\psfig{file=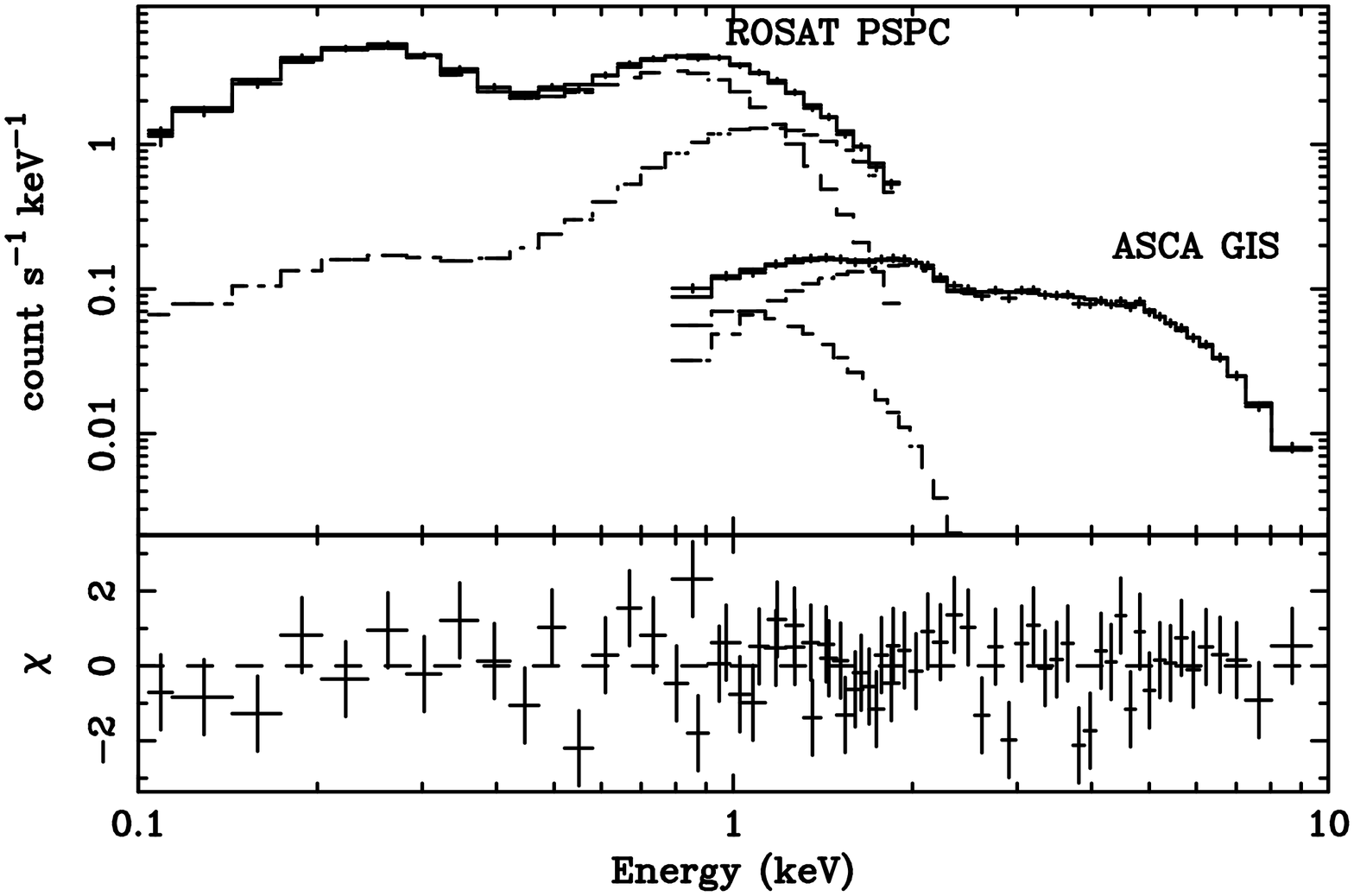,width=8cm}
\caption{Phase-averaged spectra. (Upper panel): Data are shown by
 crosses. The best-fit model (a thin-thermal plasma plus an ECUT
 power-law) of ROSAT PSPC and ASCA GIS are plotted with lines. 
(Lower panel):  Residuals from the best-fit model.}
\end{figure}

To  investigate the soft component,  we then simultaneously fitted the
ROSAT and ASCA spectra  in the 0.1--10.0~keV band with various
two-component models, in which an ECUT power-law model  was always
adopted to the hard component. 
For the additional soft component, we adopted a blackbody,  disk
blackbody, bremsstrahlung,  thin-thermal plasma (Mewe et al.\ 1985), and
broken power-law model. 
In these two-component model fits,  a common absorbing gas column with
solar abundance was adopted for both the hard and soft components.  
All models were statistically rejected at the 90\% confidence level with
wavy residuals below $\sim 1$~keV. In particular, an edge-like
structure was seen at $\sim 0.5$~keV, which may correspond to the
absorption edge of neutral oxygen atoms.  
Allowing  the oxygen abundance in the absorption column to be free,  we
obtained statistically acceptable fits only when the soft component was
either  a thin-thermal plasma or a broken power-law.  
The best-fit parameters for these models are separately listed in tables
1 and 2, while the best-fit model is shown in figure 3.  


We separately  made phase-resolved spectra from phases 0--0.5
(`on-pulse') 
and  0.5--1 (`off-pulse').  We fit these spectra with the two
accepted models, and found that only the normalization of the hard
component varies with the pulse phase.
The best-fit parameters of these models are also  listed in tables 1 and
2, while the best-fit model is shown in figure 4. 
We also extracted the ``pulsed spectrum'' by subtracting  the off-pulse
spectrum from that of on-pulse.  
We found that the pulsed spectrum is fitted with an ECUT power-law with
absorption, of 
which the best-fit parameters are  consistent with those of the hard
component of the phase-averaged spectrum.  These facts indicate that the 
pulsating X-rays contribute only to the hard component, 
consistent with the initial working hypothesis. 

\begin{figure}[tp]
\psfig{file=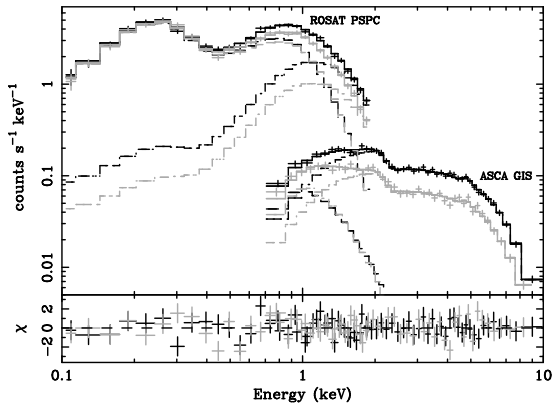,width=8cm}
\caption{Phase-resolved spectra. (Upper panel): Data are shown by
 crosses.  The  best-fit model (a thin-thermal plasma plus an ECUT
 power-law) of  `on-pulse' and  `off-pulse' are plotted with black and
 gray lines, respectively. (Lower panel):  Residuals from the best-fit
 model.}
\end{figure}

\begin{table*}[t]
\begin{center}
Table~1. \hspace{4pt}Best-fit parameters for a thin-thermal plasma plus
 an ECUT power-law model. \\
\end{center}
\vspace{6pt}

\begin{tabular*}{\textwidth}{@{\hspace{\tabcolsep}
\extracolsep{\fill}}lcccccccccc} \hline \hline \\ [-6pt]
&$kT$ &$Z^*$&
$\Gamma$&$E_{\rm c}$&$E_{\rm f}$&
 $N_{\rm H}^{\dagger}$&$Z_{\rm O}^{\ddagger}$&
$L_{\rm xs}^{\S}$&$L_{\rm xh}^{\S}$&
$\chi^2/$d.o.f.  \\ 
&(keV)&{\small ($10^{-2}$~solar)}&
&(keV)&(keV)&
($10^{20}$~cm$^{-2}$)&(solar)&
\multicolumn{2}{c}{\small($10^{38}$~erg~s$^{-1}$)}& \\ \hline
Av.
&$0.37^{+0.02}_{-0.04}$&$1.5^{+0.8}_{-0.6}$&
$0.43^{+0.05}_{-0.05}$&$6.4^{+0.8}_{-0.9} $ & $9.3^{+11.2}_{-4.3}$&
$4.2^{+0.2}_{-0.2}$&$7.3^{+1.9}_{-1.8}$& 
0.8&1.8&63.9/62 \\ 
On
&$0.35^{+0.04}_{-0.03}$&$1.6^{+1.4}_{-0.9}$&
$0.48^{+0.06}_{-0.07}$&$6.5^{+1.1}_{-1.1} $&$8.5^{+11.7}_{-4.5}$&
$4.3^{+0.3}_{-0.3}$&$6.4^{+2.8}_{-2.4}$&
0.8&2.2& 48.6/62\\ 
Off
&$0.38^{+0.03}_{-0.04}$&$1.6^{+1.4}_{-0.8}$&
$0.47^{+0.10}_{-0.10}$&$5.9^{+1.3}_{-1.0} $&$8.3^{+16.3}_{-5.1}$&
$4.0^{+0.3}_{-0.3}$&$8.7^{+3.2}_{-2.5}$&
0.8&1.2& 59.4/59\\[4pt]
 \hline
\end{tabular*}

\vspace{6pt}\par\noindent

Note---The errors are  at the 90\% confidence level. \\
$*$: Metal abundance of the thin-thermal plasma in unit of 1\% of
 the solar abundance.\\ 
$\dagger$: Absorption column density with the solar abundance (except
 oxygen) in unit of $10^{20}{\rm ~H~cm}^{-2}$.\\  
$\ddagger$: Oxygen abundance in the absorbing matter in
 unit of the solar abundance. \\
\S: Absorption corrected luminosities of the soft and hard
 components in the  0.1--10.0~keV band at a 
 distance of 60~kpc,  in unit of $10^{38}~{\rm erg~s}^{-1}$.  \\
\end{table*}

\begin{table*}[t]
 \begin{center}
Table~2. \hspace{4pt}Best-fit parameters for a broken power-law plus an
  ECUT power-law model.\\
\end{center}
\vspace{6pt}
\begin{tabular*}{\textwidth}{@{\hspace{\tabcolsep}
\extracolsep{\fill}}lccccccccccc} \hline \hline \\[-6pt]
& $\Gamma_1^*$ &$\Gamma_1'^*$&$ E_{\rm b1}^*$
&$\Gamma_2$ &$E_{\rm c}$&$E_{\rm f}$
&$N_{\rm H}^\dagger$ &$Z_{\rm O}^\ddagger $ &
$L_{\rm xs}^{\S}$&$L_{\rm xh}^{\S}$
 &{\small$\chi^2/$d.o.f.} \\
&&&(keV)&&(keV)&(keV)&{\small ($10^{20}$~cm$^{-2}$)}&
 (solar)&\multicolumn{2}{c}{\small($10^{38}$~erg~s$^{-1}$)} & 
\\[4pt]\hline\\[-6pt] 
Av.
&$2.6^{+0.3}_{-0.7}$&$5.1^{+0.9}_{-0.6}$&$0.9^{+0.1}_{-0.1}
$&$0.41^{+0.06}_{-0.07}$&$6.4^{+0.6}_{-0.9}$&$9.0^{+14.0}_{-4.0}
$&$5.0^{+0.6}_{-1.2}$&$4.3^{+2.5}_{-1.6}$ & 
0.8&1.8&63.0/61\\ 
On
&$2.5^{+0.5}_{-1.5}$ &$5.3^{+3.3}_{-0.9}$ &$0.9^{+0.2}_{-0.1}$ 
&$0.46^{+0.07}_{-0.08}$ &$6.5^{+1.0}_{-1.1}$ &$8.4^{+11.1}_{-4.4}$
&$4.9^{+1.1}_{-2.6}$ &$4.3^{+4.9}_{-2.9}$& 
0.9&2.2&48.7/61\\ 
Off
&$2.6^{+0.4}_{-0.7}$ &$5.2^{+1.8}_{-0.8}$ &$0.9^{+0.2}_{-0.1}$ 
&$0.46^{+0.12}_{-0.16}$ &$5.9^{+1.3}_{-1.0}$ &$8.1^{+16.8}_{-4.9}$ 
&$5.0^{+0.7}_{-1.4}$ &$4.5^{+2.1}_{-1.9}$& 
1.1&1.3&57.4/58\\[4pt] \hline
\end{tabular*}

\vspace{6pt}\par\noindent

Note---The errors are  at 90\% confidence level. \\
$*$: $\Gamma$ and $\Gamma'$ are photon indices below and above $E_{\rm b}$, 
respectively. \\
$\dagger$: Absorption column density with the solar abundance (except
 oxygen) in unit of $10^{20}{\rm ~H~cm}^{-2}$.\\  
$\ddagger$: Oxygen abundance of the absorbing matter in
 unit of the solar abundance. \\
$\S$: Absorption corrected luminosities of the soft and hard components
 in the  0.1--10.0~keV band at 
 a distance of 60~kpc, in unit of $10^{38}~{\rm erg~s}^{-1}$. 

\end{table*}

\section{Discussion}

\subsection{Comparison to the Previous Results}

Hughes (1994)  analyzed only the ROSAT data, and showed that 
the spectrum is fitted  by a model having an extremely soft spectrum: a
blackbody ($kT \sim 35$~eV) plus a steep  power-law ($\Gamma \sim
2.4$). 
However, we have shown that the spectrum in the ASCA band (the hard
component) is as hard as those of usual XBPs,  and the ROSAT spectrum
consists of mostly  the ``soft excess'' below $\sim 2$~keV.  

For a consistency check, we first fitted our ROSAT spectrum with Hughes' 
model, and obtained best-fit parameters consistent with Hughes (1994).
However, as can be also seen in figure 2 in Hughes (1994), systematic
positive residuals were found above $\sim 1.5$~keV, which implies
the existence of a hard component.
We then fitted the ASCA and ROSAT spectra simultaneously
with a three-component model, consisting of the 
hard ECUT power-law to the Hughes' two-component model. 
In this model, the best-fit  blackbody temperature became much lower
than the original result.  After corrections for absorption
and detector efficiency, the bolometric luminosity of the soft black
body has an unrealistic value of $\sim 10^{45}~{\rm erg~s^{-1}}$.

On the other hand, as we found, the combined ASCA and ROSAT spectra can
be fitted with 
two different components, with a more reasonable luminosity of 
$\sim 10^{38}~{\rm erg~s^{-1}}$  for both components. 
Also, the energy-dependent pulse fraction and energy-independent pulse
profile are naturally explained by our two-component model. We thus
infer that the present model is more probable, hence the spectrum of
RX~J0059.2$-$7138 is not unusual for an XBP,  at least in the hard energy
band.

\subsection{Oxygen Over-Abundance}

We found an edge-like structure at 0.5~keV, which would be a
hint of  over-abundance of oxygen in the absorbing matter.
One may argue, however, that the ROSAT spectrum fitted by Hughes (1994)
does not show any structure at 0.5~keV.  We should note that the blackbody and
power-law components in the Hughes's model  cross at around 0.5~keV,
which may produce an artificial dip, and hence would compensate for the
edge-like structure. 

To check whether the structure is real or not, we fitted a narrow-band
spectrum (0.2--0.8~keV) with  a power-law absorbed by cool matter of
solar abundance. Because the structure remained near 0.5~keV, we claim that 
the edge-like structure is real.

Since the oxygen abundance  of the SMC interstellar matter is less
than the solar value (Russell, Dopita 1992),  we can infer that there
exists oxygen-rich matter around the binary system of
RX~J0059.2$-$7138. 
The companion star, which should have ejected the  oxygen-rich matter,
may be in an evolved  stage  where the stellar surface becomes abundant
only in oxygen, although such stars (e.g.\ WO stars) are very rare.

On the other hand, metals such as carbon, nitrogen, and oxygen could be
brought to the surface of a massive star by turbulent diffusion, due to
the rapid rotation or tides in a binary system (Maeder 1987).
We thus fitted the spectra again, based on the assumption that 
abundances of carbon and nitrogen in the absorbing matter 
are the same as that of oxygen.
The abundance was then determined to be $\sim 6$ solar, 
which leads us to a more comfortable possibility that 
the surface of the companion star is enriched in CNO, which may be
caused by tidal effects in the close binary system.

We found that a thin-thermal plasma with very low abundance of 0.02 solar
can fit the soft X-ray spectrum.  As we discuss in the next subsection,
the soft component is likely to originate from a large region
surrounding the binary system. 
This creates a dilemma  that the X-ray emitting  circumstellar medium
is extremely under-abundant, while the X-ray absorbing  circumstellar gas is
CNO over-abundant.  Although we have no clear idea of how to solve this
dilemma, we suggest that the thin-thermal plasma is not a physical
model, but is  a phenomenological model like a broken power-law for the
soft component. 

\subsection{Origin of the Hard and Soft Components}

Since the pulsation was found to be mostly due to the hard component,
we can conclude that  the hard component originates from a small region,
probably near to the polar cap of the neutron star. Hence, the origin would
be the same as those of other XBPs. 
However, the cut-off energy is lower than that of the usual XBPs
(10--20~keV; e.g.\ Mihara 1995a). 
According to the correlation between the cut-off energy and magnetic field
strength (Mihara 1995a, 1995b),
the magnetic field of RX~J0059.2$-$7138 is estimated to be 
$6\times10^{11}<B<1\times10^{12}$~G, which is one of the weakest magnetic
fields in  XBPs.

A puzzle is the origin of the soft component. 
Since it shows no pulsation, 
it is likely that the soft component originates from a
relatively large region, covering a significant fraction of the full
binary system.  
In fact, when we adopt the thin-thermal plasma model as the soft component, 
the emission measure is very large: $n^2 V \sim 10^{61}~{\rm cm^{-3}}$ 
(table 1), where $n$ and $V$ are number density and volume of the plasma. 
If we assume that the plasma is distributed spherically, 
its radius should be larger than 0.07~AU 
under the optically thin condition,  
which does not conflict with the thin thermal scenario for the soft component.

At present,  only a  few XBPs  are known to have a soft excess. 
Her X-1 has a soft component which is  described by a blackbody ($kT
\sim 0.09$~keV; Dal Fiume et al.\ 1998),  and is pulsating.  McCray et
al.\ (1982) found the phase shift between pulse profiles of the soft and
the hard components,  and argues that the soft component is produced by
reprocessing of the hard component  in the inner accretion disk.  
Another XBP, LMC X-4  has a soft component of a blackbody ($kT \sim
0.03$~keV) plus a bremsstrahlung ($kT \sim 0.4$~keV) (Woo et al.\ 1996). 
The blackbody component is almost constant during  the pulse phase,
while the bremsstrahlung component is pulsating  with some phase shift
from the hard component.  
Woo et al.\ (1996) argued that  the blackbody component is emitted from
the accretion disk, and the bremsstrahlung component is from somewhere
near the neutron star, collimated as a fan-beam.  

The soft component of RX~J0059.2$-$7138 is, however, different
from these  XBPs;  it is not pulsating, it is not a blackbody spectrum
and it has no emission line.  
Hence, the origin of the soft X-rays would be different from those
proposed by the above-mentioned authors. 
As a matter of fact,  RX~J0059.2$-$7138 is a unique source among the
XBPs with soft excess; because it probably has a Be star companion (Southwell,
Charles 1996), it is different from either  low-mass companion
systems (Her X-1)  or from  high-mass supergiant
companion systems (LMC X-4). 

Broad-band spectroscopic studies on  the spectra of XBPs  with a Be star
companion have until now been very limited, due to the transient nature
of this class and the relatively narrow energy ranges of the previous and
contemporaneous satellites.  
New-generation satellites having a broad-band sensitivity and a wide
field of view, such as ASTRO-E, XMM, and Chandra Observatory, will greatly improve
our understanding on the soft excess of XBPs. 
\vspace{1pc}\par
We  express our thanks to J.\ Hughes, who told  us about the serendipitous
detection of RX~J0059.2$-$7138 in the GIS field and the referee for his
useful comments.
J.Y.\  is supported by JSPS Research Fellowship for Young
Scientists. 
This research has made use of data obtained through the High Energy
Astrophysics Science Archive Research Center Online Service,
provided by the NASA/Goddard Space Flight Center.

\small
\section*{References}
\re
 Bildsten L., Chakrabarty D., Chiu J., Finger M.H., 
 Koh D.T., Nelson R.W., Prince T.A., Rubin B.C.\ et al.\ 1997, ApJS 113,
 367 
\re
 Burke B.E., Mountain R.W., Daniels P.J., Dolat V.S., 
 Cooper M.J.\ 1994, IEEE Trans.\ Nucl.\ Sci.\ 41, 375
\re
 Dal Fiume D., Orlandini M., Cusumano G., Del Sordo S., 
 Feroci M., Frontera F., Oosterbroek T., Palazzi E.\ et al.\ 
 1998, A\&A 329, L41
\re
 Hughes J.P.\ 1994, ApJ 427, L25
\re
 Kylafis N.D.\ 1996, in Supersoft X-ray Sources, ed J.\ Greiner, 
Lecture Notes in Physics 472 (Springer, Berlin) p41
\re
 Maeder A.\ 1987, A\&A 178, 159 
\re
 Mathewson D.S.\ 1985, Proc.\ Astron.\ Soc.\ Aust.\  6, 104
\re
 McCray R.A., Shull J.M., Boynton P.E., Deeter J.E., Holt S.S., White N.E.
 1982, ApJ 262, 301
\re
 Mewe R., Gronenschild E.H.B.M., van den Oord G.H.J.\ 1985,
       A\&AS 62, 197
\re
 Mihara T.\ 1995a,  PhD Thesis, The University of Tokyo
\re
 Mihara T.\ 1995b, RIKEN Review No.\ 10
\re
 Nagase F.\ 1989, PASJ 41, 1
\re
 Ohashi T., Ebisawa K., Fukazawa Y., Hiyoshi K., 
 Horii M., Ikebe Y., Ikeda H., Inoue H.\ et al.\ 1996, PASJ 48, 157
\re
 Russell S.C., Dopita M.A.\ 1992, ApJ 384, 508
\re
 Serlemitsos P.J., Jalota L., Soong Y., Kunieda H., Tawara Y., 
 Tsusaka Y., Suzuki H., Sakima Y.\ 
 et al.\ 1995, PASJ  47, 105
\re
 Southwell K.A., Charles P.A.\ 1996, MNRAS 281, L63
\re
 Stella L., White N.E., Rosner R.\ 1982, ApJ 308, 669
\re
 Tanaka Y., Inoue H., Holt S.S.\ 1994, PASJ 46, L37
\re
 Tr\"{u}mper J.\ 1983, Adv.\ Space Sci.\ 2, 241
\re
 Woo J.W., Clark G.W., Blondin J.M., Kallman T.R., Nagase F.\ 1995, ApJ
 445, 896
\re
 Woo J.W., Clark G.W., Levine A.M., Corbet R.H.D., Nagase F.\ 1996, ApJ
 467, 811
\re
 Yokogawa J., Paul B., Ozaki M., Nagase F., Chakrabarty D., 
 Takeshima T.\ 2000, ApJ, submitted

\end{document}